\documentclass[a4paper,fleqn,usenatbib]{mnras}

\usepackage{newtxtext,newtxmath}
\usepackage[T1]{fontenc}
\usepackage{ae,aecompl}

\usepackage{graphicx}	% Including figure files
\usepackage{amsmath}	% Advanced maths commands
\usepackage{bm}

\makeatletter
\newcommand{\customlabel}[2]{%
   \protected@write \@auxout {}{\string \newlabel {#1}{{#2}{\thepage}{#2}{#1}{}} }%
   \hypertarget{#1}{#2}
}
\makeatother

\title[Solar wind temperature anisotropy-driven instabilities]{On the interplay of  solar wind proton and electron instabilities: Linear and quasi-linear approaches}

\author[S.M.Shaaban et al.]{
S. M. Shaaban,$^{1,2}$\thanks{E-mail: s.m.shaaban88@gmail.com}
M. Lazar,$^{3,4}$
R. A. L{\'o}pez,$^{5}$
R. F. Wimmer-Schweingruber$^{1}$
\\
 $^{1}$Institute of Experimental and Applied Physics, University of Kiel, Leibnizstrasse 11, D-24118 Kiel, Germany.\\
$^{2}$Theoretical Physics Research Group, Physics Department, Faculty of Science, Mansoura University, 35516, Mansoura, Egypt.\\
$^{3}$Centre for Mathematical Plasma Astrophysics, KU Leuven, Celestijnenlaan 200B, B-3001 Leuven, Belgium.\\
$^{4}$Institut f\"ur Theoretische Physik, Lehrstuhl IV: Weltraum- und Astrophysik, Ruhr-Universit\"at Bochum, D-44780 Bochum, Germany.\\
$^5$Departamento de F{\'i}sica, Universidad de Santiago de Chile, USACH, 9170124 Santiago, Chile.
}

\date{Accepted XXX. Received YYY; in original form ZZZ}

\pubyear{2020}

\begin{document}
\label{firstpage}
\pagerange{\pageref{firstpage}--\pageref{lastpage}}
\maketitle

% Abstract of the paper
\begin{abstract}
Important efforts are currently made for understanding the so-called kinetic instabilities, driven by the anisotropy of different species of plasma particles present in the solar wind and terrestrial magnetosphere. These instabilities are fast enough to efficiently convert the free energy of plasma particles into enhanced (small-scale) fluctuations with multiple implications, regulating the anisotropy of plasma particles. In this paper we use both linear and quasilinear (QL) frameworks to describe complex unstable regimes, which realistically combine different temperature anisotropies of electrons and ions (protons). Thus parameterized are various instabilities, e.g., proton and electron firehose, electromagnetic ion cyclotron, and whistler instability, showing that their main linear properties are markedly altered by the interplay of anisotropic  electrons and protons. 
Linear theory may predict a strong competition of two instabilities of different nature when their growth rates are comparable. In the QL phase wave fluctuations grow and saturate at different levels and temporal scales, by comparison to the individual excitation of the proton or electron instabilities. In addition, cumulative effects of the combined proton and electron induced fluctuations can markedly stimulate the relaxations of their temperature anisotropies. Only whistler fluctuations inhibit the efficiency of proton firehose fluctuations in the relaxation of anisotropic protons.
These results offer valuable premises for further investigations in numerical simulations, to decode the full spectrum of kinetic instabilities resulting from the interplay of anisotropic electrons and protons in space plasmas.
\end{abstract}

% Select between one and six entries from the list of approved keywords.
% Don't make up new ones.

%%%%%%%%%%%%%%%%% BODY OF PAPER %%%%%%%%%%%%%%%%%%
%_______________________
\section{Introduction}
%_______________________
%
Space plasmas are in general collision-poor if not even collision-less, and their dynamics should be governed by the interaction of plasma particles with the waves and fluctuations of electromagnetic fields. The observations confirm the existence of these fluctuations, either as a multi-scale turbulent spectrum \citep{Alexandrova2009}, or as coherent, small-scale wave fluctuations, most probably, enhanced by kinetic instabilities \citep{Wilson-etal-2013, Gary2016}. Particularly important are the instabilities driven by temperature anisotropy $A \equiv T_{\perp}/ T_{\parallel} \ne 1$ of plasma particles, where $\perp$ and $\parallel$ denote the perpendicular and parallel directions to the magnetic field. In the solar wind, anisotropic temperatures are predicted by various physical mechanisms, like magnetic compression inducing an anisotropy $A>1$, e.g., in the outer corona,  or the adiabatic expansion of solar wind along a decreasing magnetic field, leading to an opposite anisotropy~$A<1$. Self-generated instabilities (see Appendix A for a short summary) are expected to explain not only the enhanced fluctuations \citep{Bale2009,Wilson-etal-2013}, but to also regulate the increase of temperature anisotropy and explain the observations \citep{Hellinger2006, Bale2009, Michno2014, Shaaban2017}.

In the last decade important efforts have been made, in an attempt to build realistic theories, and describe not only individual instabilities, but also the interplay of different instabilities, as driven,  cumulatively, when both the electrons and protons are anisotropic \citep{Lazar2011, Michno2014, Shaaban2016, Maneva2016, Shaaban2017, Shaaban2018, Ali2020}. From a more realistic mixing of the anisotropies of electrons and protons multiple instabilities can be triggered, e.g., electron and proton firehose instabilities at similar time scales, or the electron and proton (ion) cyclotron instabilities. The interplay of these instabilities lead to new unstable regimes, which are not characterized yet beyond a linear approach. Thus, recently it was shown that for such complex cases, combining different particle populations with various anisotropies, quasi-stable states predicted by linear theory, i.e., below the instability thresholds, may not be confirmed by the extended quasi-linear approaches, which indicate a deeper relaxation of anisotropic populations after the instability saturation \citep{Shaaban2019AA,Shaaban2020}.
Moreover, recent attempts to describe the effects of anisotropic electrons on proton firehose instability (PFHI) in simulations \citep{Micera2020} have used extremely large temperature anisotropies for both electrons and protons, i.e. $A_{p,e}=0.1$, which made unstable modes intermingle such that a straightforward identification of the operative time scale of each instability was impossible, and, therefore, physical interpretations remained incomplete. 
Another recent QL analysis was carried out by \cite{Ali2020}, who examined the interplay of EM ion-cyclotron (EMIC) and electron firehose (EFH) instabilities, concluding that particles react on different spatial and temporal scales. However, the cumulative effects of these two instabilities on the relaxation of temperature anisotropies of protons and electrons remain to be investigated in depth. 

Thus motivated, in this paper we investigate  alternative regimes combining different temperature anisotropies of electrons and protons, typically for solar wind conditions. We present the most relevant results of a comprehensive  QL analysis, providing valuable insights from the saturation of instabilities via a complex relaxation of the proton and electron temperature anisotropies, corresponding to their characteristic time scales.
Our paper is organized as follows:
Section~\ref{sec:2} presents, in brief, the linear and QL approaches of parallel electromagnetic instabilities, for complex but realistic plasma conditions, considering both the electron and proton populations anisotropic. Numerical solutions are discussed for three alternative regimes of combined excitations, starting with the interplay of the proton firehose (PFH) and electron firehose (EFH) instabilities in Section~\ref{sec:3.1}, PFH and whistler instability (WI) in Section~\ref{sec:3.2}, and the EM ion-cyclotron (EMIC) and EFHI in section~\ref{sec:3.3}. Our results enable to identify the contributions of each plasma species and instability mechanisms to the combined excitation, quantify the operative time scale of each cumulative excitation, and, also, differentiate the contributions of enhanced fluctuations to the relaxation of protons and electrons. The effects of the electron temperature anisotropy $A_e>1$ on the QL evolution of EMIC instability and the associated relaxation of the proton temperature anisotropy are described in section~\ref{sec:3.4}. In section~\ref{sec:4} we summarize the results of the present analysis, and discuss their potential implications in solar wind observations.

%
%______________________________________________
\section{Linear and quasi-linear approaches}\label{sec:2}
%_____________________________________________
%
At time scales characteristic of kinetic instabilities we can consider a homogeneous and collisionless solar wind plasma of temperatures anisotropic protons (subscript $j=p$) and electrons (subscript $j=e$). For simplicity both species are described by a bi-Maxwellian distribution function in $( v_{\parallel },v_{\perp })$ velocity space
\begin{align}\label{e1}
f_{j}\left( v_{\parallel },v_{\perp }\right) =&\frac{1}{\pi
^{3/2}\alpha_{j \perp}^{2} ~ \alpha_{j \parallel}}\exp \left(
-\frac{v_{\parallel }^{2}} {\alpha_{j \parallel}^{2}}-\frac{v_{\perp
}^{2}}{\alpha_{j \perp}^{2}}\right),   
\end{align}
with thermal velocities $\alpha_{j \perp, \parallel}(t)=\sqrt{2k_B T_{j \perp, \parallel}(t)/m_j}$ (evolving in time $t$ in our QL approach) defined in terms of the temperature ($T$) components in the perpendicular ($\perp$) and parallel ($\parallel$) directions to the background magnetic field, where $m_j$ is the mass of the species~$j$.

The linear (instantaneous) dispersion relation of the electromagnetic modes propagating in directions parallel to the background magnetic field, i.e., ${\bm k}\times{\bm B}_0$=0, in the normalized form reads \citep{Shaaban2017}
\begin{align} \label{e2}
\tilde{k}^2=&~\mu~\left(A_e-1\right)+\mu~ \frac{A_e(\tilde{\omega}\mp\mu)\pm\mu}{\tilde{k} \sqrt{\mu~\beta_{e \parallel}}} Z_{e}\left(\frac{\tilde{\omega}\mp\mu}{\tilde{k} \sqrt{\mu~\beta_{e \parallel}}}\right)
\nonumber\\
&+A_p-1+\frac{A_p~\left(\tilde{\omega}\pm 1\right) \mp 1}{\tilde{k} \sqrt{\beta_p}} Z_{p}\left(\frac{\tilde{\omega}\pm 1}{\tilde{k} \sqrt{\beta_{p \parallel}}}\right)
\end{align} 
where $\tilde{k}=ck/\omega_{p p}$ is the normalized wave-number ($k$), $c$ is the speed of light, $\omega_{p p}=(4\pi n_p e^2/m_p)^{1/2}$ is the proton plasma frequency,  $\tilde{\omega}=~\omega/\Omega_p$ is normalized wave frequency ($\omega$), $\Omega_p=~e B_0/m_p c$ is the non-relativistic proton gyro-frequency, $\mu=m_p/m_e$ is the proton to electron mass ratio, $A_j\equiv \beta_{j \perp}/\beta_{j \parallel}$ and $\beta_{j \perp,\parallel}=8\pi n_a k_B T_{j \perp, \parallel }/B_0^2$, respectively, are the temperature anisotropies and plasma beta parameters for protons, and electrons, $\mp$ denote, respectively, the circular left-handed (LH) or right-handed (RH) polarization, and  
\begin{equation}  \label{e3}
Z_{j}\left( \xi _{j}^{\pm }\right) =\frac{1}{\sqrt{\pi}}\int_{-\infty
}^{\infty }\frac{\exp \left( -x^{2}\right) }{x-\xi _{j}^{\pm }}dx,\
\ \Im \left( \xi _{j}^{\pm }\right) >0, 
\end{equation}
is the standard plasma dispersion function \citep{Fried1961}.

In the quasi-linear (QL) formalism, we solve QL equations for both
particles and electromagnetic waves. The time evolution of the particle velocity distributions are characterized by the particle kinetic equation in the diffusion approximation as follows \citep{Yoon2017R}
\begin{align} \label{e4}
\frac{\partial f_j}{\partial t}&=\frac{i e^2}{4m_j^2 c^2~ v_\perp}\int_{-\infty}^{\infty} 
\frac{dk}{k}\left[ \left(\omega^\ast-k v_\parallel\right)\frac{\partial}{\partial v_\perp}+ 
k v_\perp\frac{\partial}{\partial v_\parallel}\right]\nonumber\\
&\times~\frac{ v_\perp \delta B^2(k, \omega)}{\omega-kv_\parallel-\Omega_j}\left[ 
\left(\omega-k v_\parallel\right)\frac{\partial}{\partial v_\perp}+ k v_\perp
\frac{\partial}{\partial v_\parallel}\right]f_{j}, 
\end{align}
where $f_{j}$ is the velocity distribution function for protons and electrons, and $\delta B^2(k)$ is the spectral wave energy of the enhanced fluctuations, which is described by the wave kinetic equation 
\begin{equation} \label{e5}
\frac{\partial~\delta B^2(k)}{\partial t}=2 \gamma_k \delta B^2(k),
\end{equation}
with the (instantaneous) growth rate $\gamma_k$ of the plasma instabilities calculated from the linear dispersion relation \eqref{e2}. 

The time evolution of the temperature components in the perpendicular and parallel directions $T_{j \perp,\parallel}$ (the second order moments of the particles velocity distributions) for protons ($j=p$), and electrons ($j=e$) are obtained from \eqref{e4} as follows 
\begin{subequations}\label{e6}
\begin{align}
\frac{dT_{j \perp}}{dt}&=\frac{1}{2}\frac{\partial}{\partial t}\int d{\bm v}~m_j v_\perp^2~f_j,\\
\frac{dT_{j \parallel}}{dt}&=\frac{\partial}{\partial t}\int d{\bm v}~m_j v_\parallel^2~f_j,
\end{align}
\end{subequations}

Detailed mathematical derivations of equations \eqref{e6} can be found in \cite{Seough2012, Yoon2017R, Sarfraz2017, Shaaban2019apj, Lazar2019, Ali2020}. However, for the sake of completeness here we present the QL equations after derivations in the normalized form as follows 
\begin{subequations}\label{e7}
\begin{align}
\frac{d\beta_{p \perp}}{d\tau}=&-\int\frac{d\tilde{k}}{\tilde{k}^2} W(\tilde{k})\left\lbrace\left(2 A_p-1\right)\tilde{\gamma}+\text{Im} \frac{2i\tilde{\gamma} \pm 1}{\tilde{k}\sqrt{\beta_{p \parallel}}}~\eta_p^{\pm}\right\rbrace,\\
\frac{d\beta_{p \parallel}}{d\tau}=&2\int\frac{d\tilde{k}}{\tilde{k}^2} W(\tilde{k})\left\lbrace  A_p~\tilde{\gamma}+\text{Im} \frac{\tilde{\omega} \pm 1}{\tilde{k}\sqrt{\beta_{p \parallel}}}~\eta_p^{\pm}\right\rbrace,\\
\frac{d\beta_{e \perp}}{d\tau}=&-\int\frac{d\tilde{k}}{\tilde{k}^2} W(\tilde{k})\left\lbrace \mu \left(2 A_e-1\right)\tilde{\gamma}+\text{Im} \frac{2i\tilde{\gamma} \mp \mu}{\tilde{k}\sqrt{\beta_{e \parallel}}}~\eta_e^{\mp}\right\rbrace,\\
\frac{d\beta_{e \parallel}}{d\tau}=&2\int\frac{d\tilde{k}}{\tilde{k}^2} W(\tilde{k})\left\lbrace \mu~A_e~\tilde{\gamma}+\text{Im} \frac{\tilde{\omega} \mp \mu}{\tilde{k}\sqrt{\beta_{e \parallel}}}~\eta_e^{\mp}\right\rbrace
\end{align}
\end{subequations}
with
\begin{align*}
\eta_p^{\pm}=&\left[A_p~\tilde{\omega}\pm\left(A_p-1\right)\right]Z_{p}\left(\frac{\tilde{\omega}\pm 1}{\tilde{k} \sqrt{\beta_{p \parallel}}}\right),\\
\eta_e^{\mp}=&\sqrt{\mu}\left[ A_e~\tilde{\omega}\mp\left(A_e-1\right)\mu\right]Z_{e}\left(\frac{\tilde{\omega}\mp\mu}{\tilde{k} \sqrt{\mu~\beta_{e \parallel}}}\right),
\end{align*}
and 
\begin{align}\label{e8}
\frac{\partial~W(\tilde{k})}{\partial \tau}=2~\tilde{\gamma}~ W(\tilde{k}).
\end{align}
where $W(\tilde{k})=\delta B^2(\tilde{k})/B_0^2$ is the normalized spectral wave energy, $\tau=t~\Omega_p$ and $\tilde{\gamma}=\gamma/\Omega_p$.
%
%__________________________________
\section{Stability analysis}\label{sec:3}
%__________________________________
%
This section presents the results of the numerical linear and QL analyses of the electromagnetic instabilities propagating at directions parallel to the background magnetic field ($\bm{k}\parallel \bm{B}_0$). We solve the QL equations~\eqref{e7} and~\eqref{e8} numerically for eleven different sets of initial plasma parameters, which we name cases \ref{c1}--\ref{c11} in the following. QL analysis enables us to study in details the instability development and its enhanced fluctuations as well as their reactions back on the macroscopic plasma parameters such as the plasma beta parameters $\beta_{\perp, \parallel j}$, and temperature anisotropies $A_j$. In all the numerical QL analyses discussed in the present manuscript, we adopt an initial wave noise of intensity $W(\tilde{k}, 0)=~10^{-6}$.
%
%------------------------------------------------------
\begin{figure}
\centering 
\includegraphics[width=0.47\textwidth, trim=2.6cm 7.2cm 2.cm 3.2cm, clip]{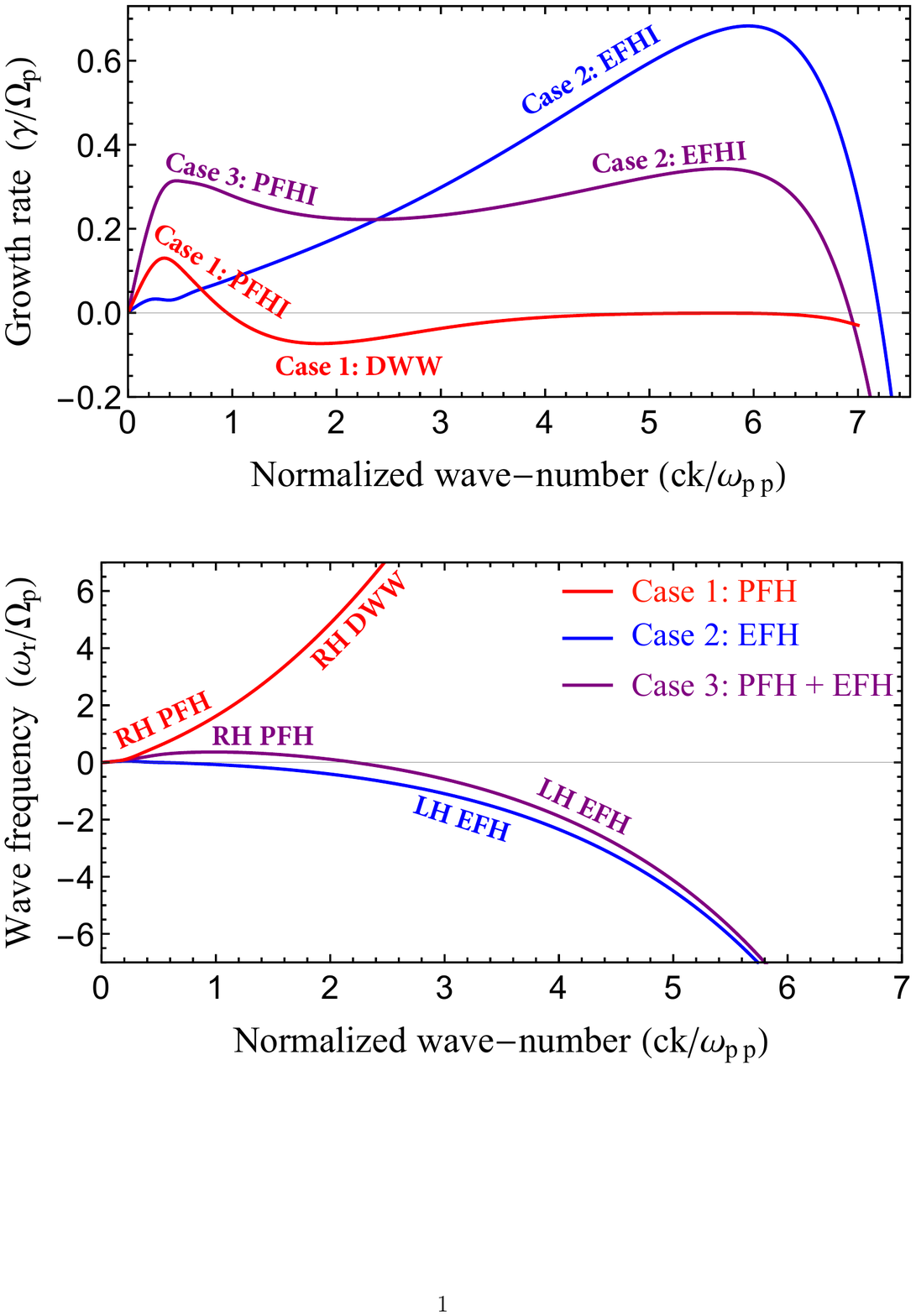}
\caption{Growth rates $\gamma/\Omega_p$ (top) and wave frequencies $\omega_r/\Omega_p$ (bottom) of the FHI driven by either anisotropic protons $A_p=0.4$ in case~\ref{c1} (red), or anisotropic electrons $A_e=0.45$ in case~\ref{c2} (blue), or cumulatively by anisotropic protons $A_p=0.4$ and anisotropic protons $A_e=0.45$ in case~\ref{c3} (purple).}
\label{f1}
\end{figure}
%------------------------------------------------------
%
%----------------------------------------------------
\begin{figure*}
\centering 
\includegraphics[width=0.96\textwidth, trim=2.6cm 17.4cm 2.cm 2.6cm, clip]{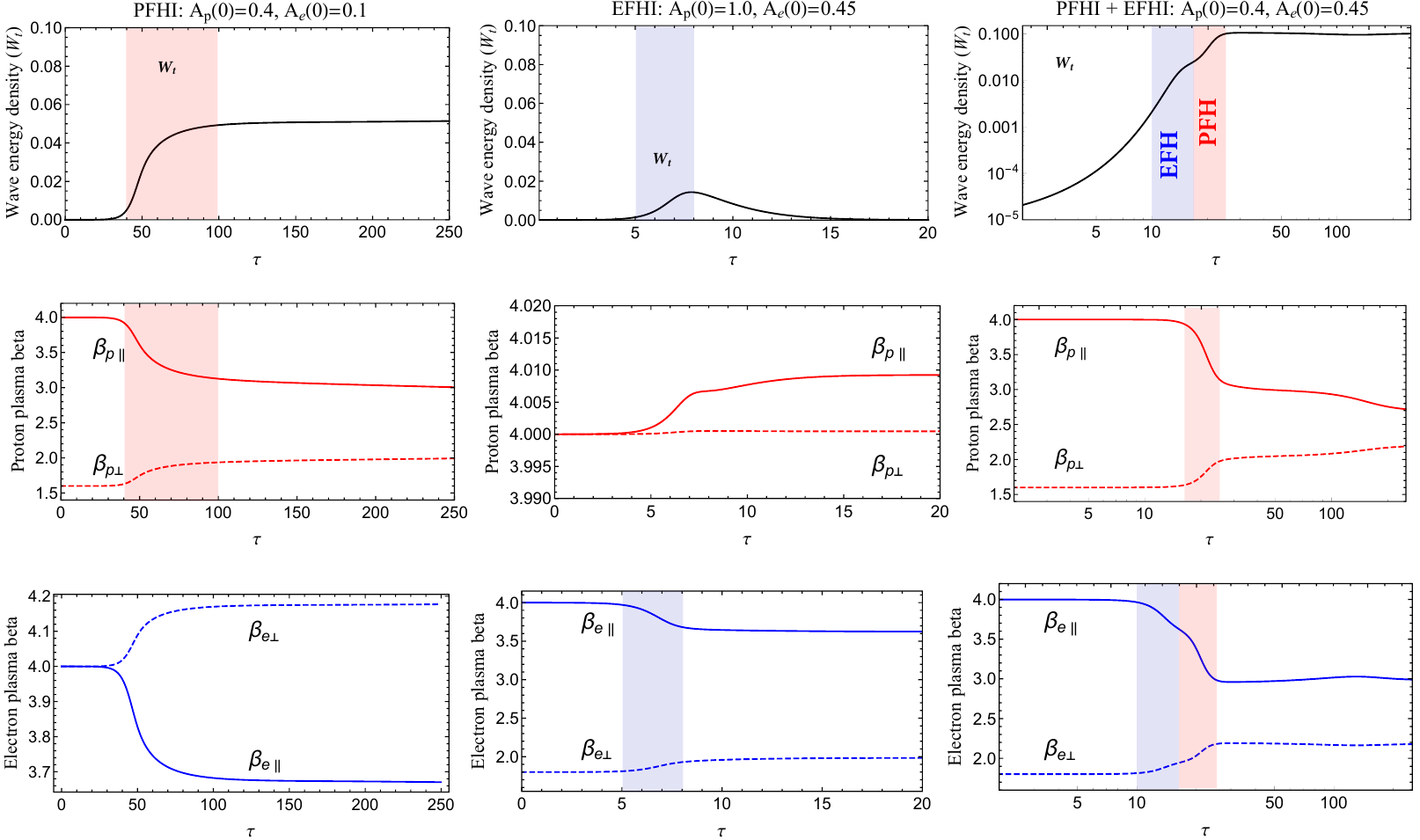}
\caption{Temporal profiles of the magnetic wave energy density $W_t$  (top), plasma betas of protons $\beta_{p \perp, \parallel}$ (middle row), and electrons $\beta_{e \perp, \parallel}$ (bottom) for case~\ref{c1} (left), \ref{c2} (middle column), and \ref{c3} (right).}
\label{f2}
\end{figure*}
%----------------------------------------------------
%

%
%---------------------------------------------------
\begin{figure}
\centering 
\includegraphics[width=0.455\textwidth, trim=2.7cm 7.6cm 2.cm 3.4cm, clip]{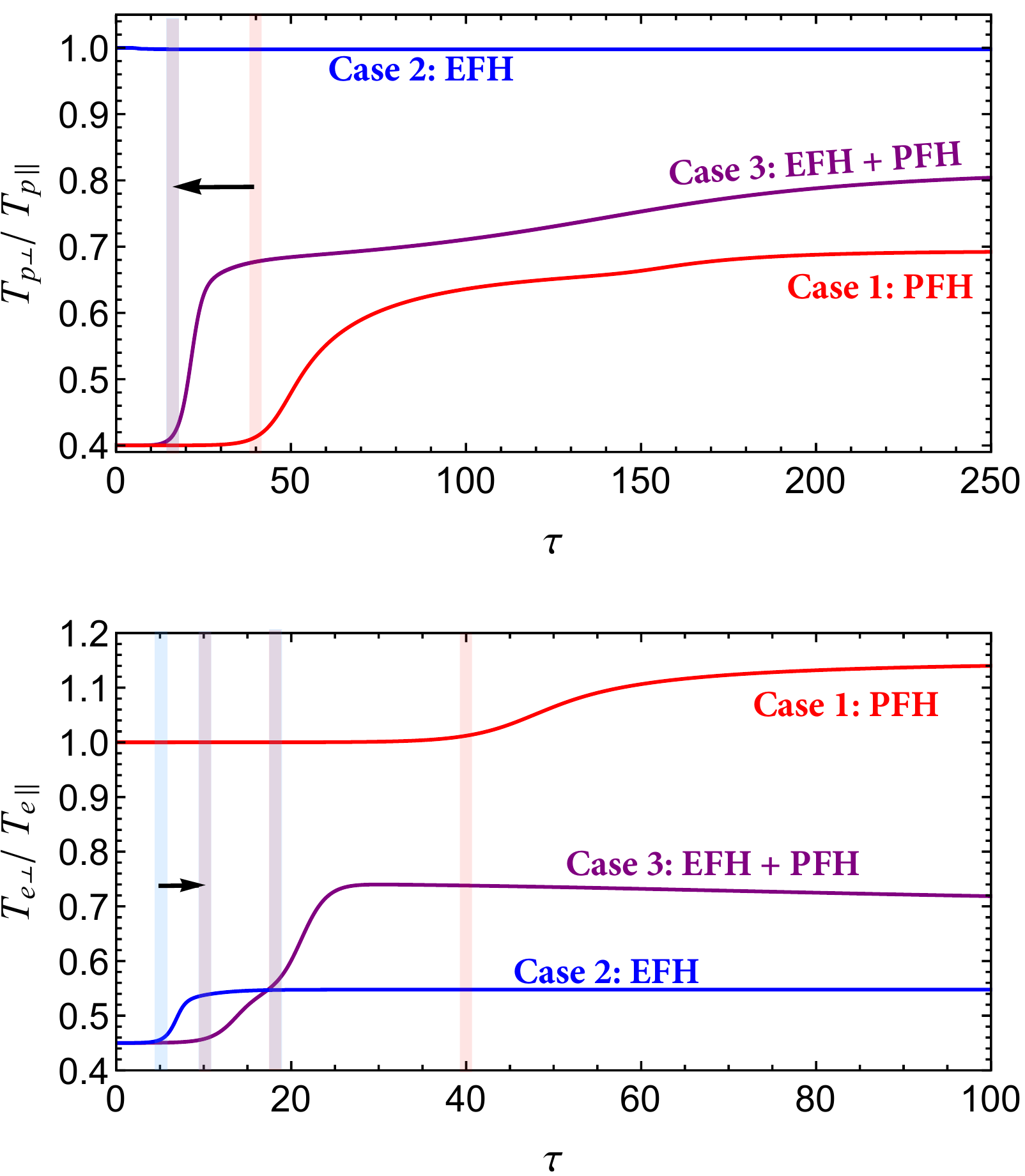}
\caption{Temporal profiles of the temperature anisotropies of protons (top) and electrons (bottom) for cases~\ref{c1} (red), \ref{c2} (blue), and ~\ref{c3} (purple).}
\label{f3}
\end{figure}
%-----------------------------------------------------
%
%____________________________________________________________
\subsection{The interplay of PFHI and EFHI}\label{sec:3.1}
%____________________________________________________________
%
\begin{itemize}
\item {Case~\customlabel{c1}{\color{blue}1}}: $A_p(0)=0.4$, and $A_e(0)=1.0$,
\item {Case~\customlabel{c2}{\color{blue}2}}: $A_p(0)=1.0$, and $A_e(0)=0.45$,
\item {Case~\customlabel{c3}{\color{blue}3}}: $A_p(0)=0.4$, and $A_e(0)=0.45$.
\end{itemize}

Other plasma parameters used in our calculations for cases~\ref{c1}--\ref{c3} are $\beta_{p \parallel}(0)=4.0$, and $\beta_{e \parallel}(0)=4.0$.  

In order to highlight the cumulative effects induced by the interplay of the solar wind proton and electron temperature anisotropies on the firehose instabilities (FHI) and the associated electromagnetic fluctuations, as well as their actions back on the particle velocity distributions we perform a comparative analysis against the idealized models, which ignore the mutual effects between electrons and protons. Thus, Figure~\ref{f1} displays the growth rates $\gamma/\Omega_p$ (top panel) and the wave frequencies $\omega_r/\Omega_p$ (bottom panel) of the FHI driven by either temperature anisotropic protons $A_p(0)=0.4$ (case \ref{c1}), or temperatures anisotropic electrons $A_e(0)=0.45$ (case \ref{c2}), or both $A_p(0)=0.4$ and $A_e(0)=0.45$ (case \ref{c3}) for plasma beta parameters $\beta_{p \parallel}(0)=\beta_{e \parallel}(0)=4.0$. The unstable solutions for cases~\ref{c1} and~\ref{c2} display two individual peaks which correspond to growth rates of PFHI (red) and EFHI (blue), respectively. For case~\ref{c3} the growth rate displays two distinct but connected peaks, the first peak at the low wavenumbers corresponds to the PFHI and the second peak at larger wave numbers corresponds to the EFHI, see the purple line. The corresponding wave frequency for case~\ref{c3} confirms the conversion of the RH polarized PFH to the LH polarized EFH modes at large wave numbers by changing the sign in between the PFHI and EFHI peaks, see the purple line. For isotropic electrons $A_e=1.0$ (case~\ref{c1}) the wave frequency of the RH polarized PFH mode at low wavenumbers increases monotonically to the electron scales describing the damped RH whistler wave (DWW) with frequency  $\omega_r>\Omega_p$, see the read line. A comparison between the growth rates shows that the PFHI is stimulated by the electron anisotropy $A_e=0.45$ (compare red and purple lines), while the EFHI is inhibited by the proton anisotropy $A_e=0.4$ (compare blue and purple lines).

Results from linear approaches can describe the steady state of the plasma instabilities, i.e., at $t=0$, but cannot describe either their evolution or their reactions back to the particles velocity distributions. Instead, it is the quasi-linear (QL) analysis that enable us to study and understand the instability evolution upon the saturation and its interaction with the plasma particles in the velocity space. Thus, in Figure~\ref{f2} we study the QL temporal profiles of the magnetic wave energy density $W_t=\int d\tilde{k}~W(\tilde{k})$ (top row) of the PFHI (left column) and EFHI (middle column) fluctuations, and their back reactions on the plasma beta parameters (or dimensionless temperatures) of protons $\beta_{p \parallel, \perp}$ (middle row), and electrons $\beta_{e \parallel, \perp}$ (bottom row) in contrast with those of the combined PFHI and EFHI fluctuations in case~\ref{c3} (right column). The QL development of PFHI (case~\ref{c1}) and EFHI (case~\ref{c2}) are well known, e.g., \cite{Seough2012, Seough2015, Sarfraz2017, Yoon2017e, Shaaban2019apj}, but the presence of these idealized cases here enables straightforward comparisons with the more realistic situation, i.e., case~\ref{c3}, in which we consider the interplay of the PFHI and EFHI fluctuations. 

Left-top panel of Figure~\ref{f2} shows that the magnetic wave energy density $W_t$ of the PFH fluctuations (case~\ref{c1}) is growing in the time interval $\tau=[40-100]$ before eventually becomes saturated, see the red shaded area. The parallel component of the proton plasma beta $\beta_{p \parallel}$ (red solid line) is strongly decreased, while the perpendicular component $\beta_{p \perp}$ (red dashed line) is increased as time progresses in the interval $\tau=[40-100]$ as a reaction of the $W_t$ enhancement, see left-middle panel. Initially isotropic electrons, i.e., $A_e(0)=1$, in the left-bottom panel are subjected to parallel ($\parallel$) cooling and perpendicular ($\perp$) heating as indicated by $\beta_{e \parallel}$ (blue solid line) and $\beta_{e \perp}$ (blue dashed line), respectively, and become temperature anisotropic in the perpendicular direction at later stages, i.e., $A_e(\tau_m)>1$. 

The middle column of Figure~\ref{f2} displays the QL development of EFHI driven by an excess of the parallel electron temperature $A_e(0)<1$ for the initial plasma parameters given in case~\ref{c2}. The top panel shows that the initial magnetic wave energy density $W_t$ increases in the time interval $\tau=[5-8]$ (see the blue shaded area) and peaks at $\tau=8$ before suffering a reabsorption by the plasma particles. As direct consequences the initially anisotropic electrons (bottom panel) are subjected to strong parallel cooling and perpendicular heating in the time interval $\tau=[5-8]$, as indicated by $\beta_{e \parallel}$ (blue solid line) and $\beta_{e \perp}$ (blue dashed line), respectively. The middle panel shows that initially isotropic protons,i.e., $A_p(0)=1.0$, experience a very week heating process in both parallel and perpendicular directions, which is increased as time progresses for $\tau>8$, suggesting that the wave energy reabsorption in the top panel is caused by protons. Similar temporal profiles for $W_t$, $\beta_{p \parallel,\perp }$, and $\beta_{e \parallel,\perp }$ associated with the excitation of the EFHI have been recently reported in \cite{Ali2020}, but for a different range of plasma parameters. A series of behaviors that are evident from left (case~\ref{c1}) and middle (case~\ref{c2}) columns of Figure~\ref{f2} may be noted as follows: The magnetic wave energy density $W_t$ of the EFH fluctuations is much faster, but less intense than that for PFH fluctuations. The QL development of the PFHI is associated with a considerable induced temperature anisotropy for electrons in the perpendicular direction after the instability saturation, i.e., $A_e(\tau_m)>1$. The effects of the EFHI development on the temporal profile of the initially isotropic protons are minimal.      

The QL results for the combined development of the PFHI and EFHI driven by the interplay of initially anisotropic protons and electrons in case~\ref{c3}, i.e., $A_p(0)=0.4$ and $A_e(0)=0.45$, are displayed in the right column of Figure~\ref{f2}. The top panel shows that magnetic wave energy $W_t$ first experiences an exponential growth in the time interval $\tau=[10-16]$ and then steps into another phase, in which $W_t$ undergoes a second exponential growth in the time interval $\tau=(16-25]$ followed by saturation. As we learned from cases~\ref{c1} and \ref{c2} that enhancement of $W_t$ for EFH fluctuations is faster, but less intense than that for PFHI, we can confirm that the first exponential growth of $W_t$ at early time corresponds to the EFHI (see the blue shaded area), and the second exponential growth at later time corresponds to the PFHI (see the red shaded area). Further confirmations can be made by studying the temporal profiles of the proton plasma beta parameters $\beta_{p \parallel, \perp}$ and $\beta_{p \parallel, \perp}$ in the middle and bottom panels, respectively. It is clear that the parallel cooling (solid lines) and perpendicular heating (dashed lines) processes on electron population (bottom panel) are much faster than those for the proton population (middle panel). A comparison between temporal profiles of $W_t$ and $\beta_{p \parallel, \perp}$ for cases \ref{c1} and \ref{c3} suggests that $W_t$ is triggered faster and reached higher level of intensity by initially temperature anisotropic electrons with $A_e(0)=0.45$. As a direct consequence the cooling and heating processes are faster and become more efficient in the relaxation of the initial proton temperature anisotropy. These results are consistent with the predictions from linear theory that show stimulating effects of the electron temperature anisotropy $A_e<1$ on the PFHI, see the purple line in Figure~\ref{f1}. A comparison between the temporal profiles of $\beta_{e \parallel, \perp}$ (bottom panels) in cases~\ref{c2} (middle column) and \ref{c3} (right column) suggests that the cooling and heating processes become slower in the presence of initially anisotropic protons $A_p(0)=0.4$. Again, this behaviour is consistent with linear theory that shows that proton temperature anisotropy $A_p(0)<1$ has inhibiting effects on the EFHI. A normal prediction that can be made from linear theory is that $A_p(0)=0.4$ can reduce the efficiency of the EFHI in the relaxation of electron anisotropy. However, in the QL phase we observe that the parallel cooling (blue solid line) and perpendicular heating (blue dashed line) processes on the electron population become more efficient in the operative regime of PFHI, i.e., for $\tau>16$, see the red shaded area in right bottom panel, a behaviour that cannot be predicted from linear theory. 

A summary for the temporal profiles of the plasma beta parameters (or dimensionless temperatures), i.e., $\beta_{p \parallel, \perp}$ (or $T_{p \parallel, \perp}$) and $\beta_{e \parallel, \perp}$ (or $T_{e \parallel, \perp}$), in Figure~\ref{f2} is displayed in Figure~\ref{f3} as temporal evolution of the temperature anisotropies of protons $A_p\equiv~T_{p \perp}/T_{p \parallel}$ (top) and electrons $A_e\equiv~T_{e \perp}/T_{e \parallel}$ (bottom). The top panel shows that the relaxation of the initial proton temperature anisotropy $A_p$ (case~\ref{c1}) is boosted by the interplay of the electron and proton temperature anisotropies and becomes $2.5$ times faster (case~\ref{c3}), as shown by light-purple and light-red lines at $\tau=16$ and $\tau=40$, respectively. The effects of the EFHI on the isotropic state of the protons (case~\ref{c2}) are negligible, see the blue line. The bottom panel suggests that the relaxation of the initial electron temperature anisotropy $A_e$ (case~\ref{c2}) is markedly boosted by the interplay of the EFHI and PFHI fluctuations but becomes two times slower (case~\ref{c3}), see the light-blue and light-purple lines at $\tau=5$, and $\tau=10$, respectively. Again the previous effect cannot be predicted from linear theory. The existing QL studies of PFHI often ignore the electron dynamics \citep{Seough2012, Seough2015}. However, the bottom panel of Figure~\ref{f3} shows that in the QL phase effects of the PFHI on the initially isotropic electron cannot be neglected as the electrons gain a considerable temperature anisotropy in the perpendicular direction at later stages, i.e., $A_e(\tau_m)>1$, see the red line. The QL relaxation profiles of the proton and electron anisotropies in Figure~\ref{f3} are in qualitative agreement with those obtained recently from PIC simulations for different plasma conditions \citep[see figure 5 therein]{Micera2020}. 
%
%______________________________________________
\subsection{The interplay of PFHI and Whistler instability}\label{sec:3.2}
%______________________________________________
%
%-------------------------------------------------------
\begin{figure}
\centering 
\includegraphics[width=0.46\textwidth, trim=2.6cm 7.2cm 2.cm 3.2cm, clip]{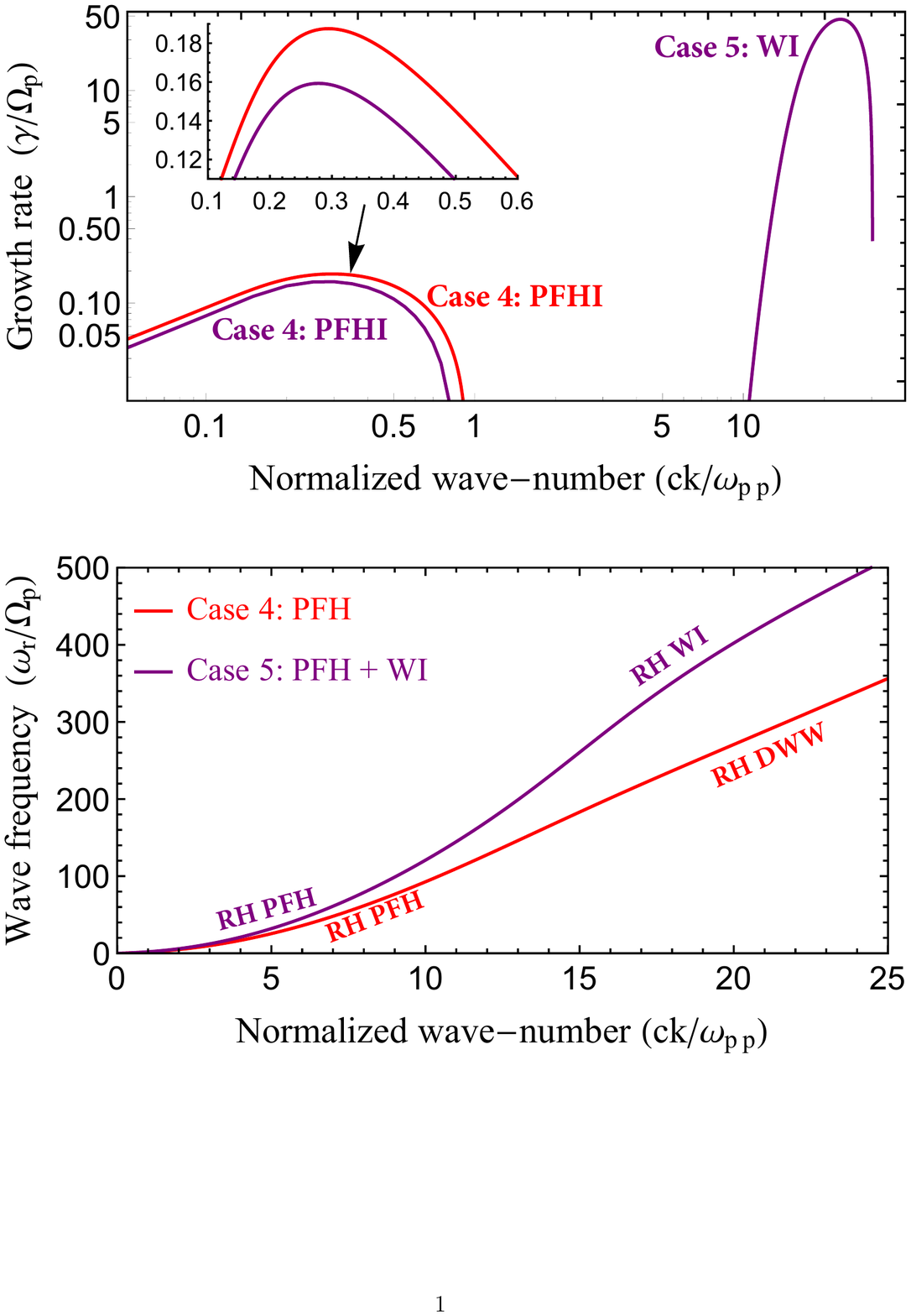}
\caption{Growth rates (top) and wave frequencies (bottom) of the PFHI driven by $A_p=0.4$, $A_e=1.0$ in case~\ref{c4} (red), PFHI and WI driven by $A_p=0.4$, $A_e=1.5$ in case \ref{c5} (purple). Other plasma parameters are $\beta_{p \parallel}=6$, and $\beta_{e \parallel}=1.0$.}
\label{f4}
\end{figure}
%-------------------------------------------------------
%
%-------------------------------------------------------
\begin{figure*}
\centering 
\includegraphics[width=0.68\textwidth, trim=3.2cm 13.7cm 2.6cm 2.6cm, clip]{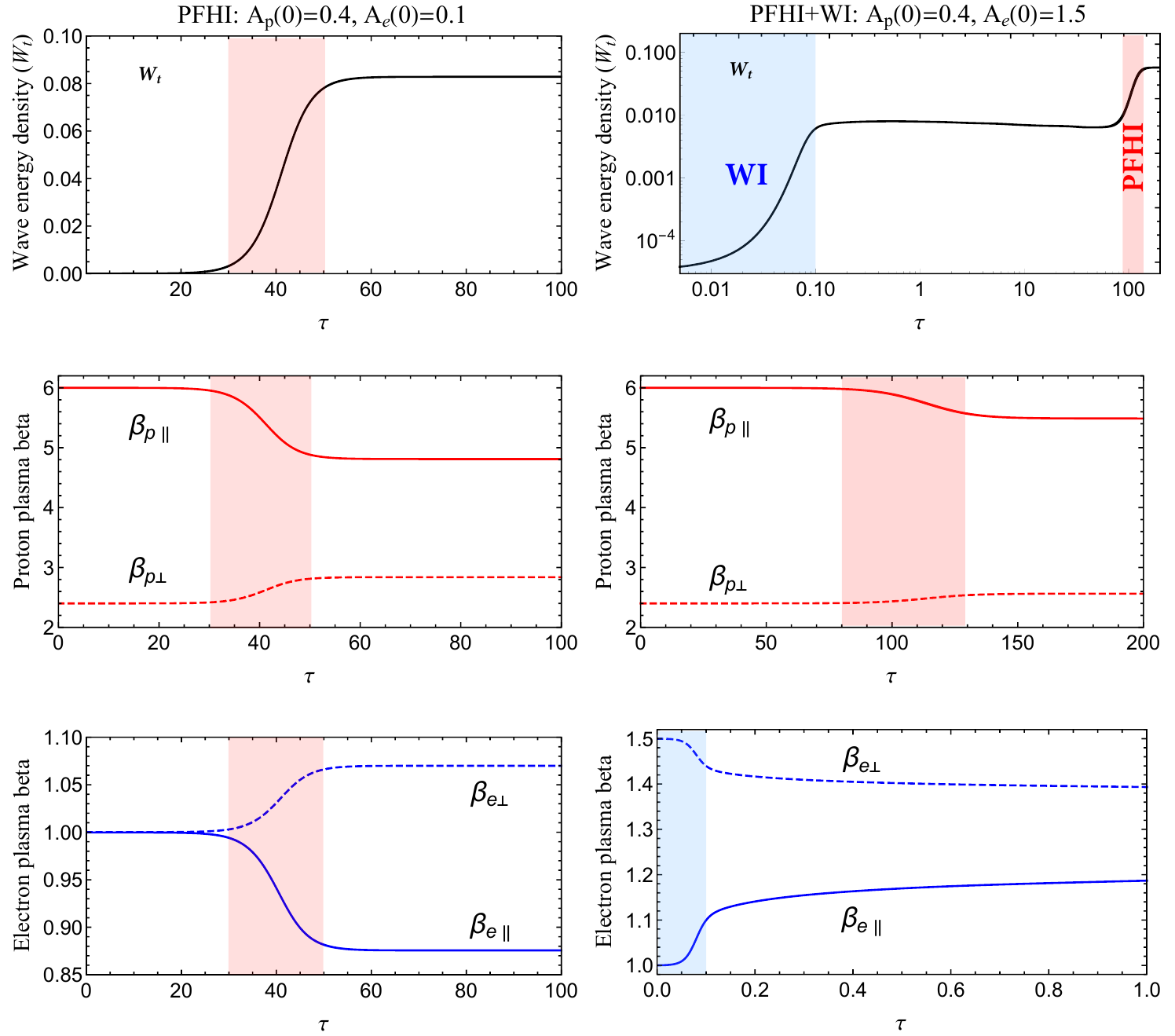}
\caption{Temporal profiles of the magnetic wave energy density $W_t$ (top), plasma betas of protons $\beta_{p \perp, \parallel}$ (middle), and electrons $\beta_{e \perp, \parallel}$ (bottom) for case~\ref{c4} (left), and case~\ref{c5} (right).}
\label{f5}
\end{figure*}
%------------------------------------------------------------
%
%-------------------------------------------------------------
\begin{figure}
\centering 
\includegraphics[width=0.46\textwidth, trim=2.6cm 7.8cm 2.cm 3.2cm, clip]{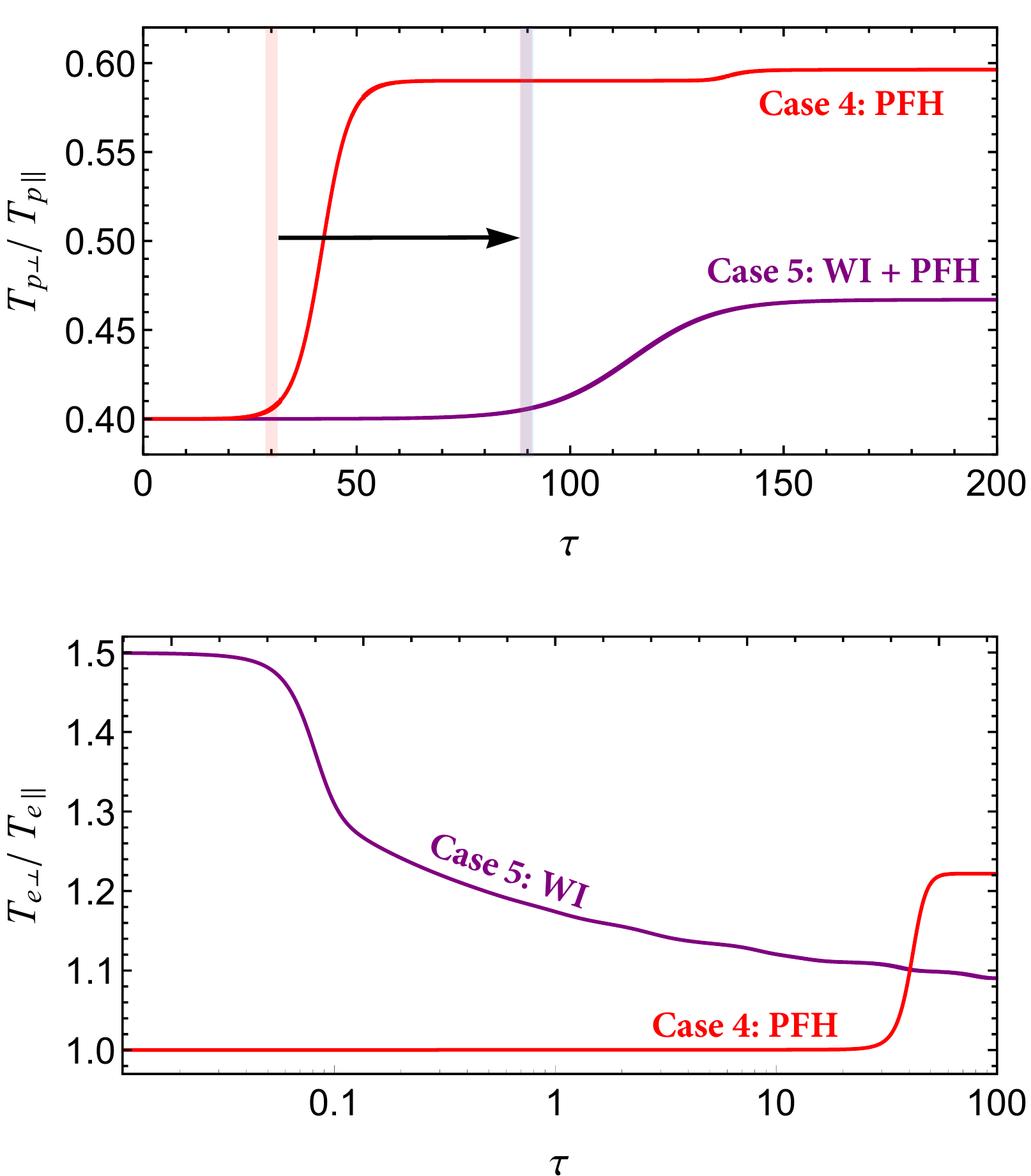}
\caption{Temporal profiles of the temperature anisotropies of protons (top) and electrons (bottom) for cases \ref{c4} (red), and \ref{c5} (purple).}
\label{f6}
\end{figure}
%------------------------------------------------------------
%
In this section we study the effects introduced by the combined excitation of PFHI and whistler instability (WI) on the temporal profiles of the macroscopic plasma parameters, i.e., $\beta_{p \perp, \parallel}$, $\beta_{e\perp, \parallel}$, and $A_{p,e}$. In order to highlight these effects we perform a comparative analysis between the idealized case that focuses only on the PFHI driven by $A_p(0)=0.4$ with $A_e(0)=1$, which we name case~\ref{c4}, and the combined excitation of PFHI and WI by proton and electron anisotropies $A_p(0)=0.4$ and $A_e(0)=1.5$, which we named case~\ref{c5}. 
\begin{itemize}
\item {Case~\customlabel{c4}{\color{blue}4}}: $A_p(0)=0.4$, and $A_e(0)=1.0$,
\item {Case~\customlabel{c5}{\color{blue}5}}: $A_p(0)=0.4$, and $A_e(0)=1.5$.
\end{itemize}

Other plasma parameters used in our calculations for cases~\ref{c4} and~\ref{c5} are $\beta_{p \parallel}(0)=6$, and $\beta_{e \parallel}(0)=1.0$.  

Figure~\ref{f4} presents the growth rates $\gamma/\Omega_p$ (top) and wave frequencies $\omega_r/\Omega_p$ (bottom) of the PFHI (red) driven by protons with $A_p(0)=0.4$ (case~\ref{c4}), and the combined PFHI and WI (purple) driven by the interplay of the proton and electron temperature anisotropies (case~\ref{c5}), i.e., $A_p(0)=0.4$ and $A_e(0)=1.5$. For case~\ref{c4} (red line) the growth rate displays a single peak corresponding to the PFHI, while for case~\ref{c5} (purple line) the growth rate display two distinct peaks, the first peak at low wavenumbers corresponds to PFHI and the second peak at larger wavenumbers corresponds to WI. Electron temperature anisotropy $A_e>1$ has inhibiting effects on the PFHI, decreasing the growth rate and the unstable wavenumbers, see the zoomed subplot. The effect of $A_e$ on the PFH wave frequency is minimal. The corresponding wave frequencies are RH polarized for both PFH and whistler modes, and at large wavenumbers, i.e., $ck/\omega_{pp}>1$, the wave frequencies increase monotonically to the electron scales describing the DWW with $\omega_r>\Omega_p$. It is also worth to noting that the wave frequency is markedly increased when whistler modes become unstable, see the purple line. 

Beyond the linear theory, in Figure~\ref{f5} we study the QL temporal evolution of the magnetic wave energy density $W_t$ (top), and plasma beta parameters for protons $\beta_{p \perp,\parallel}$ (middle) and electrons $\beta_{e \perp,\parallel}$ (bottom) for case~\ref{c5} (right column) in contrast with the idealized situation in case~\ref{c4} (left column). For case~\ref{c4} the temporal profiles of the macroscopic plasma parameters, i.e., $W_t$, $\beta_{p \perp,\parallel}$, and $\beta_{e \perp,\parallel}$, are similar to those in case~\ref{c1} despite the different initial conditions. However, for the sake of comparison between cases~\ref{c4} and~\ref{c5}, it is important to mention that for case~\ref{c4} the wave energy density $W_t$ undergoes an exponential growth in the time interval $\tau=[30-50]$, and in response protons and electrons experience strong parallel cooling and perpendicular heating in the same time interval, see the red shaded areas in the left column. 

For case~\ref{c5} the top-right panel shows the temporal profile of the magnetic wave energy density $W_t$ which experiences an exponential rise in the time interval $\tau=(0-0.1]$, peaking at $\tau=0.1$ and then saturates before stepping into another regime, in which $W_t$ undergoes a second exponential growth in the time interval $\tau=[80-130]$ followed by another saturation. Temporal profiles of plasma beta parameters in the middle and bottom panels can provide information about the nature of the enhanced fluctuations in the top panel. Electrons are immediately subjected to strong perpendicular cooling and parallel heating (normally associated with the enhanced whistler fluctuations) in the time interval $\tau=(0-0.1]$ as indicated by $\beta_{e \perp}$ (blue dashed line) and $\beta_{e \parallel}$ (blue solid line), respectively, see blue shaded area. On the other hand, protons experience strong parallel cooling and perpendicular heating (normally associated with enhanced PFH fluctuations) in the time interval $\tau=[80-130]$ as indicated by $\beta_{e \parallel}$ (red solid line) and $\beta_{e \perp}$ (red dashed line), respectively, see the red shaded area. These heating and cooling processes are consistent with the exponential growths of $W_t$ in the top panel. Thus, one can conclude that the first exponential growth at very early times corresponds to WI and the second exponential growth at later times corresponds to PFHI. A comparison between the temporal profiles of $\beta_{p \parallel, \perp}$ in cases~\ref{c4} (middle left) and~\ref{c5} (middle right) suggests that the interplay of the PFHI and WI reduce the efficiency of the cooling and heating processes on the proton population. Moreover, these processes are delayed in time in agreement with the predictions from linear theory that show an inhibition of EMIC in the presence of temperatures anisotropic electrons with $A_e(0)=1.5$. 

Figure~\ref{f6} displays a summary of the temporal profiles of $\beta_{p \perp, \parallel}$ and $\beta_{e \perp, \parallel}$ shown in Figure~\ref{f5}, as temporal evolution of the temperature anisotropies of protons $A_p=T_{p \perp}/T_{p \parallel}$ (top) and electrons $A_e=T_{e \perp}/T_{e \parallel}$ (bottom), respectively. In the top panel it is clear that the interplay of PFHI and WI reduced markedly the efficiency of the PFHI fluctuations in the relaxation of the proton temperature anisotropy $A_p$, see the purple line. The relaxation of $A_p$ becomes three times slower and insignificant, see the light-red and light-purple lines at $\tau=30$ and $\tau=90$, respectively. Bottom panel shows that the initial temperature anisotropy of electrons $A_e(0)=1.5$ is already relaxed to small values in a response to the enhanced WI fluctuations before stepping into the PFHI operative regime at $\tau>80$, see the purple line. However, initially isotropic electrons $A_e(0)=1$ gain an induced temperature anisotropy in the perpendicular direction at later stages, i.e., $A_e(\tau_m)>1$, in a response to the PFHI fluctuations at $\tau>30$, see the red line.   
%
%___________________________________________
\subsection{The interplay of EMIC and EFHI}\label{sec:3.3}
%___________________________________________
%
Recently, \cite{Ali2020} studied the QL combined development of EMIC and EFHI to identify the operative regime of each instability and their consequences on the temporal profiles of the macroscopic plasma parameters. However, the cumulative effects of the combined excitation of EMIC and EFHI on the relaxations of the proton and electron temperature anisotropies were not discussed in depth. Thus, in this section we revisit the interplay of the enhanced EMIC and EFHI fluctuations and their reactions back on the initial proton and electron temperature anisotropies. For a straightforward comparison we consider three distinct cases with plasma parameters dedicated to the excitation of EMIC by $A_p(0)=2.5$ (case~\ref{c6}), EFHI by $A_e(0)=0.55$ (case~\ref{c7}), and combined EMIC and EFHI cumulatively by $A_p(0)=2.5$ and $A_e(0)=0.55$ (case~\ref{c8}). 
%
%------------------------------------------------------
\begin{itemize}
\item {Case~\customlabel{c6}{\color{blue}6}}: $A_p(0)=2.5$, and $A_e(0)=1.0$,
\item {Case~\customlabel{c7}{\color{blue}7}}: $A_p(0)=1.0$, and $A_e(0)=0.55$,
\item {Case~\customlabel{c8}{\color{blue}8}}: $A_p(0)=2.5$, and $A_e(0)=0.55$,
\end{itemize}
%--------------------------------------------------------
%

Other plasma parameters used in our calculations for cases~\ref{c6}, \ref{c7}, and~\ref{c8} are $\beta_{p \parallel}(0)=1.0$, and $\beta_{e \parallel}(0)=4.0$.
%
%----------------------------------------------------------
\begin{figure}
\centering 
\includegraphics[width=0.46\textwidth, trim=2.6cm 7.2cm 2.cm 3.2cm, clip]{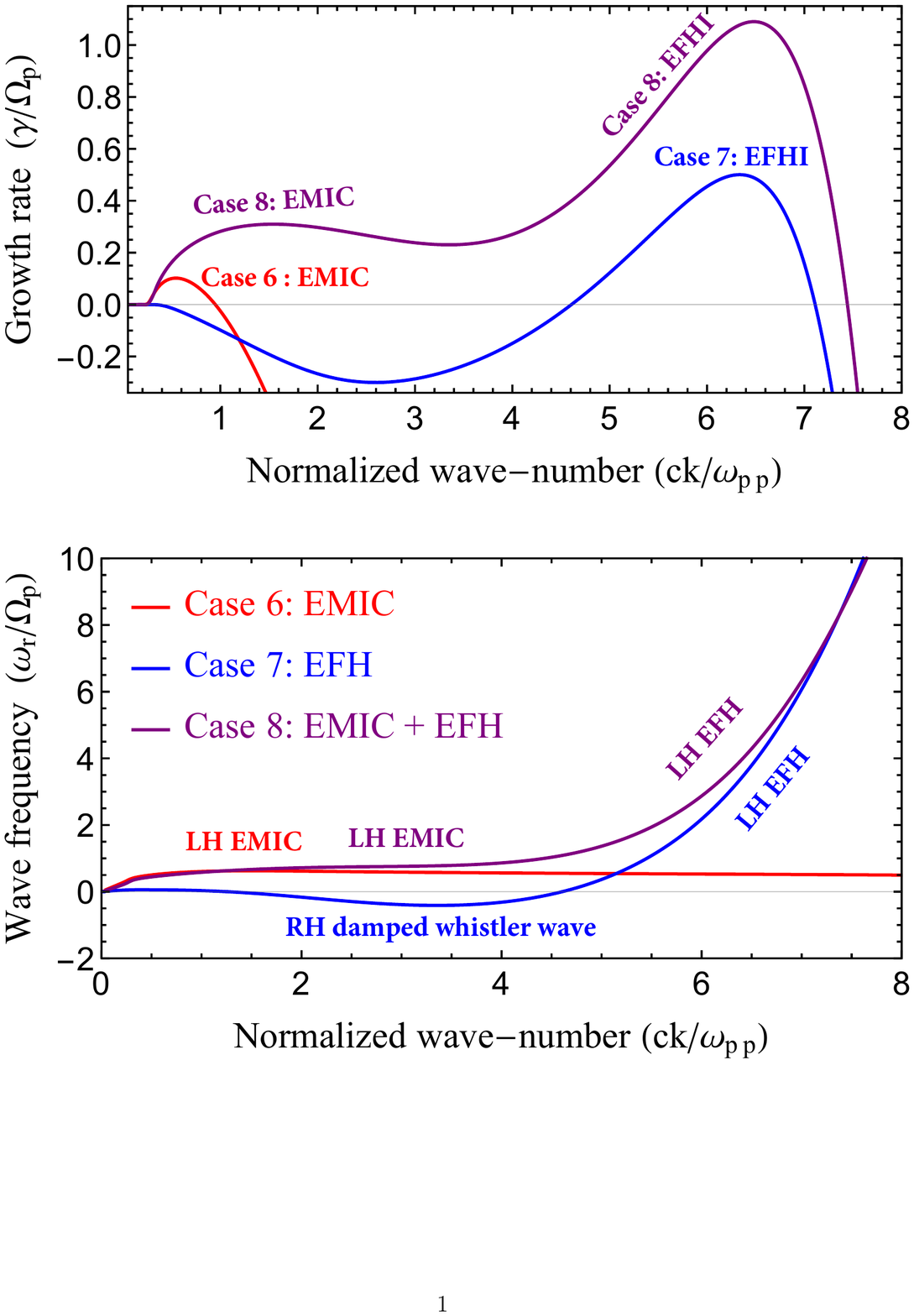}
\caption{Growth rates (top) and wave frequencies (bottom) of the EMIC driven by $A_p=2.5$ in case~\ref{c6} (red), EFHI driven by $A_e=0.55$ in case~\ref{c7} (blue), and combined EMIC and EFHI cumulatively by $A_p=2.5$ and $A_e=0.55$ in case~\ref{c8} (purple).}
\label{f7}
\end{figure}
%-------------------------------------------------------
%

%
%-------------------------------------------------------
\begin{figure}
\centering 
\includegraphics[width=0.46\textwidth, trim=4.2cm 3.4cm 3.6cm 3.2cm, clip]{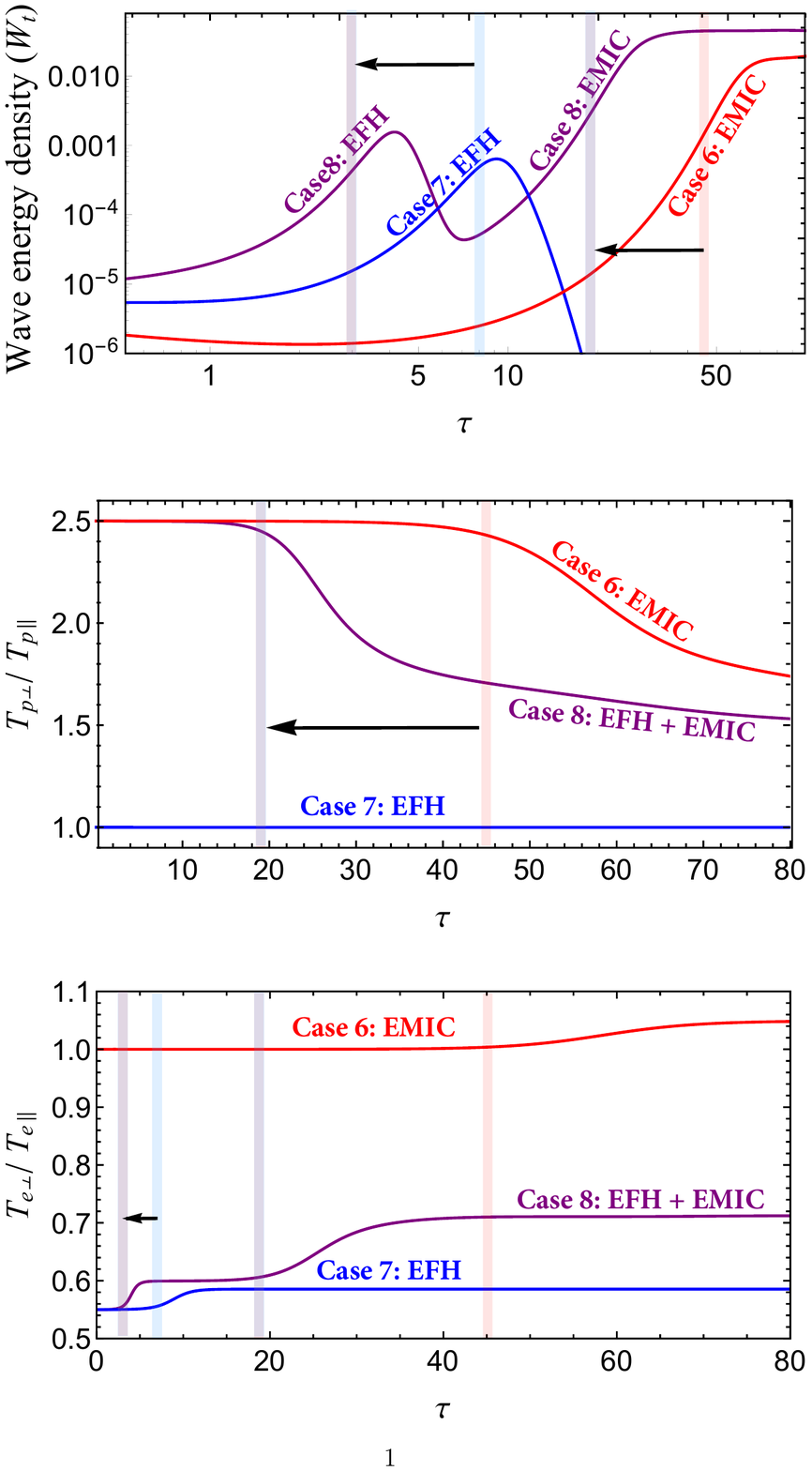}
\caption{Temporal profiles of magnetic wave energy $W_t$ (top), the temperature anisotropies of protons $A_p$ (middle) and electrons $A_e$ (bottom) for cases \ref{c6} (red), \ref{c7} (blue), and \ref{c8} (purple).}
\label{f8}
\end{figure}
%--------------------------------------------------------
%

Figure~\ref{f7} presents a comparison between the growth rates $\gamma/\Omega_p$ (top) and wave frequencies $\omega_r/\Omega_p$ (bottom) of the plasma instabilities obtained from the linear theory for the plasma parameters given in cases~\ref{c6} (red), \ref{c7} (blue), and \ref{c8} (purple). Driven by $A_p=2.5$ EMIC instability grows at small wavenumbers (red line), while EFHI driven by $A_e=0.55$ and grows at larger wavenumbers after evolving out of the RH DWW (blue line). The unstable solutions from the interplay of the proton and electron temperature anisotropies in case~\ref{c8}, i.e., $A_p=2.5$ and $A_e=0.55$, display two distinct peaks, the first peak at low wavenumbers for EMIC and the second peak at larger wavenumbers for EFHI, see the purple line. The growth rates of both EMIC and EFHI are enhanced by the interplay of the proton and electron temperature anisotropies. The cumulative effects of the proton and electrons on the wave frequencies are minimal. However, it is worth noting that in case~\ref{c8} the wave frequency is LH polarized in the wavenumbers range corresponding to EMIC instability and extended smoothly to the electron scale describing the LH polarized EFHI. 

Figure~\ref{f8} displays comparisons for the QL temporal profiles of the magnetic wave energy density $W_t$ (top), and temperature anisotropies of protons $A_p$ (middle) and electrons $A_e$ (bottom) obtained for plasma parameters given in cases~\ref{c6} (red), \ref{c7} (blue), and \ref{c8} (purple). The over-plotted vertical lines indicated the times at which the temperature anisotropies start to relax under the effect of the enhanced fluctuations, i.e., at $\tau=3$ and $8$ for EFHI, and $\tau=19$ and $45$ for EMIC instability. From these comparisons we can state the following: The magnetic wave energy density $W_t$ grows $\sim 2.3$ times faster and reaches higher level of saturation for the combined excitation of EFH and EMIC modes in case~\ref{c8} (purple line) than those of the individual excitation of EMIC in case~\ref{c6} (red) and EFHI in case~\ref{c7} (blue). As a direct consequence the relaxation of the proton temperature anisotropy $A_p$ (middle panel) becomes deeper and $\sim 2.3$ times faster in the case of the combined EMIC and EFHI (purple line) than that for the individual EMIC (red line). It is clear that the relaxation of $A_p$ enters the saturation phase in case of the combined EMIC and EFHI even before the beginning of its relaxation in the case of the individual excitation of EMIC. The effect of the EFHI on the temporal profile of the initially isotropic protons $A_p(0)=1$ (case~\ref{c7}) is negligible, see the blue line in the middle panel. On the other hand, the effects of the EMIC on the temporal profile of the initially isotropic electrons $A_e(0)=1$ (case~\ref{c6}) cannot be neglected as the electrons gain induced temperature anisotropy in the perpendicular direction at later stages, i.e., $A_e(\tau>45)>1$ in the operative regime of the EMIC fluctuations, see the red line in the bottom panel. Bottom panel shows that the relaxation of the electron temperature anisotropy $A_e$ in the case of the combined EMIC and EFHI fluctuations (purple line) is larger and also $2.3$ times faster than that for the individual excitation of EFHI fluctuations (blue line). Again, the relaxation of $A_e$ starts to saturate in case of the combined EMIC and EFHI fluctuations even before the beginning of its relaxation in the case of the individual EFHI excitation, see the light-purple and light-blue lines at $\tau=3$ and 7. It it clear that the relaxation of $A_e$ in the case of the combined EMIC and EFHI fluctuations occurs in two distinct phases, in the first phase the relaxation of $A_e$ is a response of the enhanced EFHI fluctuations in the time interval $\tau=[3,7]$ followed by saturation in the time interval $\tau=(7-19)$ before stepping into a second strong relaxation phase in a response to the enhanced EMIC fluctuations for $\tau>19$, see the purple line. These phases of relaxation for $A_e$ (bottom) are consistent with the temporal evolution of $W_t$ in the top panel (purple line), which shows that $W_t$ undergoes two distinct exponential growths corresponding to EFHI and EMIC at $\tau=3$ and 19, respectively, see the light-purple lines in top and bottom panels. 
%
%--------------------------------------------------------
\begin{figure}
\centering 
\includegraphics[width=0.46\textwidth, trim=2.6cm 7.6cm 2.cm 3.4cm, clip]{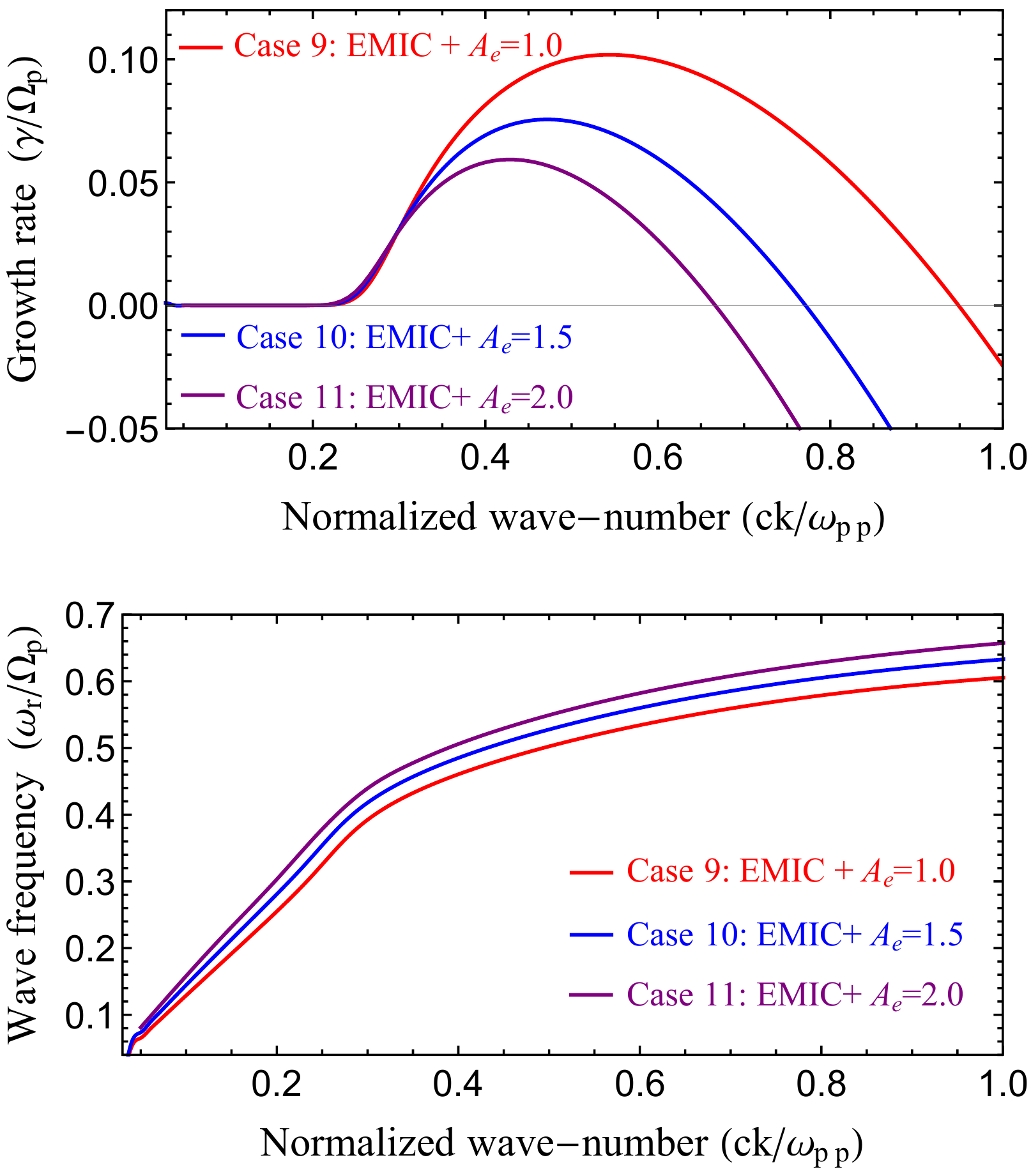}
\caption{Effects of the electron temperature anisotropy, i.e., $A_e=1.0$ (red), 1.5 (blue), and 2.0 (purple), on the  growth rates (top) and wave frequencies (bottom) of EMIC instability driven by proton anisotropy $A_p=2.5$ for $\beta_p=1.0$, and $\beta_e=2.0$.}
\label{f9}
\end{figure}
%------------------------------------------------------
%
%
%_______________________________________
\subsection{Effects of $A_e>1$ on EMIC}\label{sec:3.4}
%______________________________________
%
Recently \cite{Shaaban2017} have demonstrated that the LH EMIC and RH WI cannot interact. LH low-frequency EMIC modes cannot connect to the high-frequency whistler modes, which have opposite (RH) polarization. Thus, in this section we study the QL evolution of EMIC instability under the influence of the electron temperature anisotropy in the perpendicular direction, i.e., for $A_e(0)=1.0$ (case~\ref{c9}), $A_e(0)=1.5$ (case~\ref{c10}), and $A_e(0)=2$ (case~\ref{c11}).    
\begin{itemize}
\item {Case~\customlabel{c9}{\color{blue}9}}: $A_p(0)=2.5$, and $A_e(0)=1.0$,
\item {Case~\customlabel{c10}{\color{blue}10}}: $A_p(0)=2.5$, and $A_e(0)=1.5$,
\item {Case~\customlabel{c11}{\color{blue}11}}: $A_p(0)=2.5$ and $A_e(0)=2.0$.
\end{itemize}

Other plasma parameters used in our calculations for cases~\ref{c9}, \ref{c10}, and~\ref{c11} are $\beta_{p \parallel}(0)=1.0$, and $\beta_{e \parallel}(0)=2.0$.

%--------------------------------------------------
\begin{figure}
\centering 
\includegraphics[width=0.46\textwidth, trim=4.2cm 3.2cm 3.6cm 3.2cm, clip]{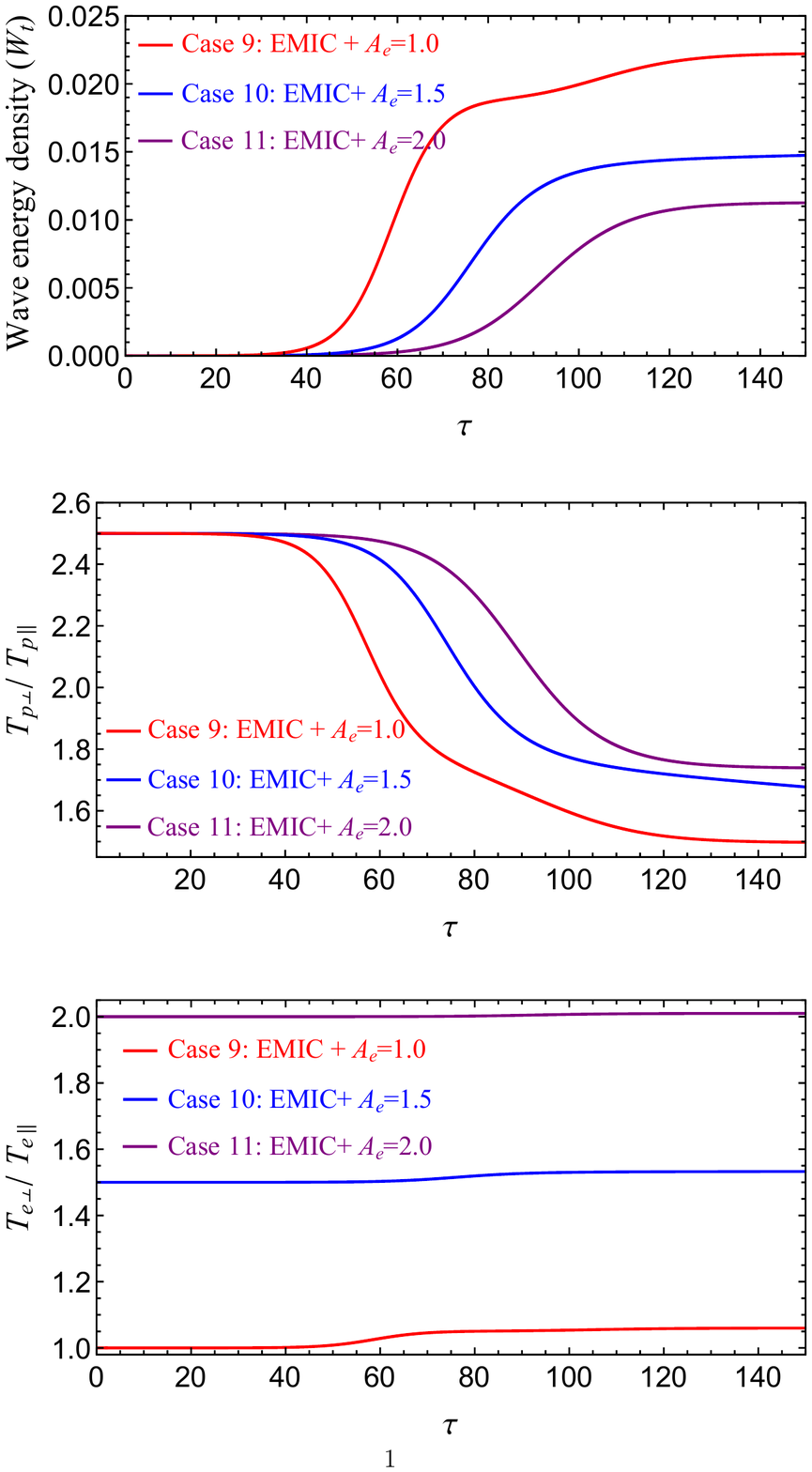}
\caption{Temporal profiles of the temperature anisotropies of protons (top) and electrons (bottom) for cases \ref{c9} (red), \ref{c10} (blue), and \ref{c11} (purple).}
\label{f10}
\end{figure}
%--------------------------------------------------
%

Figure~\ref{f9} displays the growth rates $\gamma/\Omega_p$ (top) and wave frequencies $\omega_r/\Omega_p$ (bottom) of EMIC instability driven by $A_p=2.5$ under the influence of different electron temperature anisotropies $A_e(0)=1$ (red), 1.5 (blue), and 2.0 (purple), which we name cases~\ref{c9}, \ref{c10}, and \ref{c11}, respectively. Electron temperature anisotropy $A_e>1$ has inhibiting effects on the EMIC instability, decreasing the growth rates and the unstable wavenumbers as $A_e>1$ increases, i.e., $A_e=2$. The corresponding wave frequencies are slightly increasing as $A_e>1$ increases, see the purple line. 
%
%
%%%%%%%%%%%%%%%%   Table:1   %%%%%%%%%%%%%%%%%%
\begin{table*}
	\centering
	\caption{Summary of the EM instabilities in the present paper and their cumulative effects on $\gamma$, $W_t$, $\tau$, and the relaxation of $A_j(0)$.}
   \label{t1}
	\begin{tabular}{ccccccccc} % four columns, alignment for each
		\hline
		Interplay of  & Frequency & Anisotropy & Effects on proton instabilities & Effects on electron instability\\
		\hline
		  &   &  &  Stimulate the instability: increase $\gamma$ & Inhibit the instability: decrease $\gamma$\\
		PFHI    &  RH and $\omega_r<\Omega_p$  & $A_{p}(0)<1$  &  Stimulate $W_t$ & Inhibit $W_t$\\
		EFHI    &  LH and $\omega_r>\Omega_p$  & $A_{e}(0)<1$  & Speed up the relaxation: 2.5 times faster & Slow down the relaxation: 2 times slower \\
		(Case~\ref{c3})     &  &  &  Stimulate the relaxation of $A_p(0)$ &  Markedly stimulate the relaxation of $A_e(0)$ \\
		     \hline
	            &   &  &  Inhibit the instability: decrease $\gamma$ & Negligible effects on $\gamma$\\
		PFHI    &  RH and $\omega_r<\Omega_p$  & $A_{p}(0)<1$&  Inhibit $W_t$ & Negligible effects on $W_t$\\
		WI      &  RH and $\Omega_p\ll\omega_r<|\Omega_e|$  & $A_{e}(0)>1$  & Slow down the relaxation: 3 times slower & Negligible effects on relaxation time\\
		(Case~\ref{c5})    &  &  &  Markedly inhibit the relaxation of $A_p(0)$ &  Negligible effects on $A_e(0)$ relaxation \\
		\hline
	       &   &  &  Stimulate the instability: increase $\gamma$ & Stimulate the instability: increase $\gamma$\\
		EMIC    &  LH and $\omega_r<\Omega_p$  & $A_{p}(0)>1$  &  Stimulate $W_t$ & Stimulate $W_t$\\
		EFHI    &  LH and $\omega_r>\Omega_p$  & $A_{e}(0)<1$  & Speed up the relaxation: 2.3 times faster & Speed up the relaxation: 2.3 times faster \\
		(Case~\ref{c8}) &  &  &  Stimulate the relaxation of $A_p(0)$ &  Markedly stimulate the relaxation of $A_e(0)$ \\

		\hline
		   &   &  &  Inhibit the instability: decrease $\gamma$ & Low-frequency LH EMIC modes\\
		EMIC    &  LH and $\omega_r<\Omega_p$  & $A_{p}(0)>1$  &  Inhibit $W_t$ & and high-frequency RH WI modes \\
		---------        &    ---------                          & $A_{e}(0)=2.0$  & Slow down the relaxation: 1.6 times slower  & cannot interact \\
		(Case~\ref{c11})     &  &  &  Inhibit the relaxation of $A_p(0)$ &   \\
		\hline
	\end{tabular}
\end{table*}
%%%%%%%%%%%%%%%%%%%%%%%%%%%%%%%%%%%%%%%%%
%
%

Beyond linear theory, Figure~\ref{f10} shows the QL temporal evolution of the magnetic wave energy density $W_t$ (top), temperature anisotropies of protons $A_p$ (middle) and electrons $A_e$ (bottom) for case~\ref{c9} (red), \ref{c10} (blue), and \ref{c11} (purple). It is clear that $W_t$ shows systematic diminution with a delay in the initiation, a longer growing time, and markedly lower saturation levels reached in the presence of temperature anisotropic electrons in the perpendicular direction (i.e., cases~\ref{c10}, and \ref{c11}), see blue and purple lines. As a direct consequence the relaxation of $A_p$ (middle) becomes $\sim 1.6$ times slower and less efficient, confirming the predictions from linear theory that $A_e>1$ has inhibiting effects on EMIC instability, see the purple line for $A_e=2.0$. Electrons in general gain an additional induced temperature anisotropy in the perpendicular direction at later stages (bottom), i.e., $A_e(\tau_m)>A_e(0)>1$. However, these additional induced temperature anisotropies are decreased as $A_e(0)>1$ increases (purple), as a result of the reduction in $W_t$ of the EMIC fluctuations.
%
%___________________________
\section{Conclusions} \label{sec:4}
%__________________________

In this manuscript we have shown how complex can be the interplay of temperature anisotropy instabilities when both electrons and protons are assumed anisotropic. These refined models mixing electron and proton scales are not only realistic, but enable direct comparisons with idealized kinetic approaches, which ignore the interplay of different sources of free energy in the plasma system. If linear dispersion properties of wave frequencies and growth rates only predict an interplay between instabilities of different nature, the QL approaches enable for a long-term analysis up to the saturation of these instabilities, and for quantifying their contributions to the relaxation of anisotropic populations. 

Section~\ref{sec:3.1} describes the combined excitation of the PFHI and EFHI by the interplay of anisotropic protons and electrons with $A_{p,e}<1$. Growth rates (Figure~\ref{f1}) of PFHI are stimulated by the presence of the anisotropic electrons ($A_e<1$), while the growth rates of EFHI are inhibited by the presence of the anisotropic protons ($A_p<1$). In the QL phase, the magnetic wave energy $W$ undergoes two distinct exponential growths on different time scales, i.e., electron time scale corresponds to the excitation of EFHI and the proton time scale corresponds to the excitation of PFHI, see blue and red shaded areas in left-top panel of Figure~\ref{f2}. The PFHI fluctuations are stimulated in the presence of $A_e < 1$. The corresponding relaxations of the anisotropic protons becomes 2.5 times faster, see Figures~\ref{f2}--(left panels) and \ref{f3}--(top panel). In contrast, the interplay of anisotropic electrons and protons inhibits the EFHI fluctuations and slows down the relaxation of anisotropic electrons, but it becomes more effective reaching quasistable states closer to isotropy (see Figure~ \ref{f3}, bottom panel). Such cumulative effects cannot be predicted from linear theory, but are unveiled by the extended QL approaches \citep{Shaaban2019HF, Shaaban2020}.

In section~\ref{sec:3.2} we have studied the combined excitation of the PFHI, driven by anisotropic protons with $A_p<1$, and WI induced by  electrons with $A_e>1$. Growth rates of PFHI are inhibited by the anisotropic electrons (Figure~\ref{f4}), and so is the magnetic wave energy in  Figures~\ref{f5} and \ref{f6}. The corresponding relaxation of $A_p$ becomes less effective. Linear theory does not predict any effect of anisotropic protons in WI  \citep{Gary1993, Lazar2018}. Confirmations from the QL theory are presented in Figures~\ref{f5} and \ref{f6}, showing a smooth (exponential) growth of the wave energy of WI fluctuations with saturation within an short (electron) time scale, well before the starting of second growth of the PFHI fluctuations. This interval also corresponds to the (partial) relaxation of electrons ($\beta_{e \perp, \parallel}$ and $A_e$) under the effect of the WI fluctuations. Thus, the effects of anisotropic protons with $A_p<1$ on WI remain also negligible in QL and non-linear phases. 

In section~\ref{sec:3.3} we have re-examined the combined excitation of EMIC and EFHI from the interplay of anisotropic protons with $A_p>1$ and anisotropic electrons with $A_e<1$. Recently \cite{Ali2020} have shown that the EFH and EMIC operate on distinct time scales, and we have further investigated their cumulative reactions back on the relaxations of the temperature anisotropies. The magnetic wave energy undergoes two distinct growths on two distinct time scales, first corresponding to the excitation of EFHI, and then to PFHI at later time, see the purple line in Figure~\ref{f8}, top panel. In this case the interplay of anisotropic protons and electrons stimulates both the EMIC and EFHI fluctuations, and, implicitly, the relaxations of both species become more effective, faster in time and to quasistable states closer to isotropy, see middle and bottom panels of Figures~\ref{f8}. Noticeable is the fast relaxation of the electron temperature anisotropy, well before the excitation of EMIC fluctuations, see Figure~\ref{f8}, bottom panel. Again, this is another cumulative effect that cannot be predicted by linear theory.
Section~\ref{sec:3.4} describes the effects of the temperatures anisotropic electrons with $A_e>1$ on the QL development of EMIC instability. The growth rates and the unstable wavenumbers are reduced by the anisotropic electrons (Figure~\ref{f9}), which are also found to slow down the initiation of the EMIC instability and decrease the saturation levels of the magnetic wave energy. The corresponding relaxations of the proton temperature anisotropy $A_p$ becomes less effective and slower in time. 

To conclude, the interplay of temperatures anisotropy instabilities driven cumulatively by the anisotropic protons and electrons have  important consequences not only on the linear properties of the unstable spectra, but also on the time evolution of the concurrent unstable modes and on the relaxation of anisotropic populations. In many of these cases the electromagnetic instabilities are in general stimulated by these cumulative effects, but there are also situations when unstable modes are inhibited. Comparing to the results predicted by the idealized approaches, which ignore the mutual effects of anisotropic protons and electrons, our present investigations demonstrate the importance of QL approaches, which can advance beyond the limitation of linear theory, to provide a realistic description of kinetic instabilities and their implications. Table~\ref{t1} summarizes the main dispersive characteristics of these instabilities, including our new results. The refined kinetic approaches proposed in the present manuscript may be considered as an important progress towards a more realistic interpretation of the interplay of the kinetic instabilities, as well as their (cumulative) effects on the relaxations of proton and electron temperature anisotropies in space plasmas.
%
%___________________________
\section*{Acknowledgements}
%____________________________
%
The authors acknowledge support from the Katholieke Universiteit Leuven, Ruhr-University Bochum and Christian-Albrechts-Universit\"at Kiel. These results were obtained in the framework of the projects SCHL 201/35-1 (DFG-German Research Foundation), GOA/2015-014 (KU Leuven), G0A2316N (FWO-Vlaanderen). S.M.Shaaban acknowledges the Alexander-von-Humboldt Research Fellowship, Germany. R.A.L would like to thank the support of ANID Chile through FONDECyT grant No.~11201048.

%______________________________________________________
\appendix
\section{Instabilities driven by temperature anisotropy in the solar wind}
%______________________________________________________
%
For anisotropic temperatures with an excess in direction perpendicular to the magnetic field, i.e., $T_{j,\perp}>T_{j,\parallel}$, the linear theory predicts two distinct instabilities, the electromagnetic cyclotron and the aperiodic mirror instabilities \citep{Gary1992,Gary2006, Shaaban2017, Shaaban2018}. In the solar wind conditions the cyclotron modes grow faster than mirror modes, and one can deal with the electromagnetic ion cyclotron instability (EMICI) driven by the anisotropic protons, and whistler instability (WI), also known as electromagnetic electron cyclotron instability, induced by the anisotropic electrons. 
Both the EMICI and WI exhibit maximum growth rates when propagating parallel to the magnetic field ($\bm{k}\parallel\bm{B}_0$). EMICI is a left-handed (LH) circularly polarized mode with wave frequency less than the proton gyrofrequency $\omega_r > \Omega_p$, while WI is a right-handed (RH) circularly polarized mode with wave frequency much higher than the proton gyrofrequency and less than the electron gyrofrequency, i.e., $\Omega_p \ll \omega_r < |\Omega_e|$ \citep{Gary1993}. \cite{Shaaban2017} have shown that the low-frequency EMIC modes do not  interplay with the high-frequency WI.

In the opposite situation, an excess of temperature in parallel direction, i.e., $T_{j\perp} < T_{j\parallel}$, may trigger the so-called firehose instabilities, if the parallel plasma betas are sufficiently large, i.e., $\beta_{j \parallel} > 1$ \citep{Gary1993, Gary2003, Hellinger2006, Michno2014, Shaaban2017, Shaaban2019MNFH, Lopez2019}. The periodic firehose with dominant growth rate for parallel propagation  ($\bm{k}\times \bm{B}_0=0$) is RH circularly polarized if driven by anisotropic protons and LH if driven by anisotropic electrons. The second branch is purely aperiodic, with zero wave frequency ($\omega_r=0$), and develops only for oblique propagation ( $\bm{k}\times \bm{B}_0\neq0$). In general, the periodic proton firehose instability (PPFHI) grows faster than the aperiodic modes \citep{Hellinger2006, Bale2009, Michno2014, Shaaban2017}. On the other hand, the aperiodic electron firehose instability develops much faster than the periodic electron firehose instability (EFHI) \citep{Gary2003, Camporeale2008, Hellinger2014, Shaaban2019MNFH, Lopez2019}. 

Marginal stability thresholds predicted by the linear theory are represented as inverse correlation laws of the anisotropy $A_s$ as a function of parallel plasma beta $\beta_{s \parallel}$, for each species of sort $s$. Numerous studies show that these thresholds may shape, more or less, the limits of the temperature anisotropies reported by the solar wind observations \citep{Hellinger2006, Maruca2012, Lazar2017MN, Shaaban2017, Shaaban2019AA, Shaaban2019apj}. Moreover, enhanced  electromagnetic fluctuations have been observed in association with anisotropic temperatures of electrons and protons,  suggesting that these fluctuations are generated or enhanced locally by the kinetic instabilities \citep{Bale2009, Gary2016}.

%%%%%%%%%%%%%%%%%%%%%%%%%%%%%%%%%%%%%%%%%%%%%%%%%%
%___________________________
\section*{Data Availability}
%____________________________
%
The data that support the findings of this study are available from the corresponding author upon a reasonable request.
%

%%%%%%%%%%%%%%%%%%%% REFERENCES %%%%%%%%%%%%%%%%%%

% The best way to enter references is to use BibTeX:

\bibliographystyle{mnras}
\bibliography{refs} % if your bibtex file is called example.bib

%%%%%%%%%%%%%%%%%%%%%%%%%%%%%%%%%%%%%%%%%%%%%%%%%%

% Don't change these lines
\bsp	% typesetting comment
\label{lastpage}
\end{document}